\documentclass[12pt]{iopart}
\usepackage[utf8]{inputenc}
\usepackage{graphicx}
\usepackage{dcolumn}
\usepackage{bm}
\usepackage{caption}
\usepackage{amssymb}
\usepackage{multirow}

\DeclareTextFontCommand{\mytexttt}{\ttfamily\hyphenchar\font=45\relax}
 \usepackage{xcolor}
 \colorlet{BLUE}{blue}
 \colorlet{RED}{red}

\begin{document}

\title{Noncommutativity in Effective Loop Quantum Cosmology}

\author{Abraham Espinoza-Garc\'ia}
\address{Unidad Profesional Interdisciplinaria de Ingenier\'ia Campus Guanajuato del Instituto Polit\'ecnico Nacional.\\
Av. Mineral de Valenciana \#200, Col. Fraccionamiento Industrial Puerto Interior, C.P. 36275, Silao de la Victoria, Guanajuato, M\'exico.}
\ead{aespinoza@ipn.mx}

\author{Efra\'in Torres-Lomas}
\address{Departamento de F\'isica, Divisi\'on de Ciencias e Ingenier\'ias,
    Campus Le\'on, Universidad de Guanajuato.\\ C.P. 37150, Le\'on,
    M\'exico}
\ead{efrain@fisica.ugto.mx}

\author{Sinuh\'e P\'erez-Pay\'an}
\address{Unidad Profesional Interdisciplinaria de Ingenier\'ia Campus Guanajuato del Instituto Polit\'ecnico Nacional.\\
Av. Mineral de Valenciana \#200, Col. Fraccionamiento Industrial Puerto Interior, C.P. 36275, Silao de la Victoria, Guanajuato, M\'exico.}
\ead{saperezp@ipn.mx}

\author{Luis Rey D\'iaz-Barr\'on}
\address{Unidad Profesional Interdisciplinaria de Ingenier\'ia Campus Guanajuato del Instituto Polit\'ecnico Nacional.\\
Av. Mineral de Valenciana \#200, Col. Fraccionamiento Industrial Puerto Interior, C.P. 36275, Silao de la Victoria, Guanajuato, M\'exico.}
\ead{lrdiaz@ipn.mx}

\begin{abstract}
We construct a noncommutative extension of the Loop Quantum Cosmology effective scheme for the open FLRW model with a free scalar field via a theta deformation. Firstly, a deformation is implemented in the configuration sector, among the holonomy variable and the matter degree of freedom. We show that this type of noncommutativity retain, to some degree, key features of the Loop Quantum Cosmology paradigm for a free field. Secondly, a deformation is implemented in the momentum
sector, among the momentum associated to the holonomy variable and the momentum associated to the matter field. We show that the density, as in the case of Loop Quantum Cosmology is also bounded, furthermore, its maximum value is the same.
\end{abstract}


\section{Introduction}
\setcounter{page}{3}
\pagenumbering{arabic}
The idea of noncommutativity in spacetime is not new (1947) \cite{snyder}. It
was proposed as an attempt to regularize Quantum Field Theory before the
renormalization program was established. Due to its non-local behavior,
noncommutativity was quickly forgotten after renormalization proved to be
successful. In the 1970's M. Flato and co-workers proposed an alternative path
to quantization \cite{flato}, in which a deformation of the Poisson structure of
classical phase space is performed and is encoded in the moyal $\star$-product
\cite{moyal}, and generalizations of it (for a review see \cite{dito}).  In the
early 1980's mathematicians led by A. Connes succeeded in formulating what they
called Noncommutative Geometry \cite{connes}, motivated by generalizating a
classic theorem characterizing $C^{*}$-algebras.

At the end of the last century the noncommutative paradigm was resurrected, mainly due to results in String Theory \cite{connes-douglas, seiberg-witten}, in which Yang-Mills theories in a noncommutative space arise in different circumstances as effective theories when taking certain limits, for instance: the low energy limit. This renewed interest has led to a deeper understanding, from the physical and mathematical points of view, of noncommutative field theory (for a review see \cite{douglas}).
Additionally, it is believed that in a full quantum theory of gravity the continuum picture of spacetime would no longer be consistent at distances comparable to the Planck length $\ell_{p}\sim10^{-35}\mathrm{cm}$, a quantization of spacetime itself could be in order. 

A possible way to model these effects could be via an uncertainty relation for the spacetime coordinates of the form
\begin{equation}
[x^{i},x^{j}]=i\theta^{ij}.
\end{equation}
This commutation relation is the starting point for noncommutative field theories. Attempts to implement this idea for the gravitational field have led to different proposals for a noncommutative theory of gravity (see, for example, \cite{mofat}). In particular, a noncommutative generalization of Loop Quantum Gravity (LQG) was constructed in \cite{martin}. Different incarnations of noncommutative gravity have a common feature, they are highly non-linear theories, which makes them very difficult to work with.

A possible way to study noncommutativity effects in the early universe was proposed by Garc\'ia-Compe\'an et al \cite{hugo}. They implemented noncommutativity in configuration space, in contrast to noncommutative spacetime, but only after a symmetry reduction of spacetime had been imposed and the Wheeler-DeWitt quantization had been carried, giving rise to a noncommutative quantum cosmology, mathematically similar to the noncommutative quantum mechanics constructed in \cite{mezin}. Later, G. D. Barbosa and N. Pinto-Neto \cite{barbosa} introduced this minisuperspace noncommutativity already at the classical level. The idea is that, perhaps, this effective noncommutativity could incorporate novel effects and insights of a full quantum theory of the gravitational field, alongside with providing a simple framework for studying the implications of such possible noncommutative effects in the early universe through cosmological models. Since the publication of these two seminal investigations, some works along this line have been conducted. For instance, the noncommutativity of the FLRW cosmology has been studied, as well as some of the Bianchi Class A models \cite{socorro}. Quantum black holes have also been investigated within this framework \cite{julio}.

On the other hand, LQG \cite{ashtekar-lqg, thiemann} is an attempt to quantize the gravitational field taking seriously the lessons from General Relativity, that is, it aims at a full non-perturbative background independent quantization of General Relativity. It could be said that LQG is an improvement of the canonical Weeler-DeWitt quantization programme. On the other hand, Loop Quantum Cosmology (LQC) \cite{bojowald-lrr, ashtekar-lqc} is the quantization of cosmological, symmetry reduced, models following closely the ideas and methods of LQG. In this way, the LQC of the FLRW and some of the Bianchi Class A models in the presence of a massless scalar field, employed as internal time, have been constructed \cite{ashtekar-frw, corichi-closed, vandersloot, chiou, ewing-bi, ewing-bii, ewing-bix}. In particular, as a result of the underlying quantum geometry, it has been shown that the loop quantization of the FLRW models features a bounce which enables the resolution of the cosmological singularity \cite{bojowald}. The LQC of the inhomogeneous Gowdy model has also been constructed \cite{martin-benito}. 
 
As a consequence of loop quantization, the Wheeler-DeWitt equation is no longer a differential equation, but a difference equation, which is difficult to work with even in the simplest models. In order to extract physics, effective equations based on a geometrical formulation of Quantum Mechanics have been employed to study the outcome of loop quantum corrections in cosmological models \cite{ashtekar-gfqm}. For instance, the effective description of the FLRW models reproduces very well the behavior of the corresponding full loop quantization of such models.

Recently, works that focus on the possible relation, at different physical/mathematical levels, among noncommutativity and LQG have been conducted. For instance, an emergent noncommutativity in LQG is found in \cite{SAM} when constructing position operators for spin-networks; a relation between LQG-inspired deformations of spacetime and $\kappa$-Minkowski noncommutative spacetime is established in \cite{amelino-camelia}.

The present investigation aims at constructing a noncommutative effective scheme for the open FLRW model in the presence of a free scalar field, and establishing whether such noncommutative scheme could be compatible with the LQC  paradigm, in the sense of retaining key features of the LQC of the FLRW model. Along this lines, our work could be related to a minisuperspace approximation of a more fundamental noncommutative construction based on the LQG approach, such as the one in \cite{martin}, signaling possible restrictions on the way noncommutativity is to be incorporated in such general frameworks. While being related, our philosophy is different from the class of works cited in the above paragraph since we implement a simple type of noncommutativity in an LQG-related model, rather than trying to obtain noncommutativity from LQG or LQG-inspired models.

The manuscript is organized as follows:
In section II we introduce the loop variables for the open FLRW model with a free standard scalar field and the corresponding effective scheme. In section III we recall a way in which noncommutativity can be implemented in Classical Mechanics through a deformation of the product. 
Section IV is devoted to construct a noncommutative model for the effective Loop Quantum Cosmology of the open FLRW model with the help of the method introduced in section III, along the lines of \cite{barbosa}. In the remaining part of this section we recall the canonical formulation of the open FLRW model in metric variables, in the presence of a free scalar field.

\subsection{Open FLRW Model: Metric Variables}
\hfill \break
The line element of a spatially homogeneous and isotropic universe is
\begin{equation}
  ds^2= -N^2(t) dt^2 + a^2(t) \left[ \frac{dr^2}{1-\kappa r^2} + r^2
    d\Omega^2 \right] \, , \label{frw}
\end{equation}
where $a(t)$ is the scale factor, $N(t)$ the lapse function and $\kappa$ is the curvature constant, which can take values $0$, $1$ y $-1$, which correspondes to a flat, closed and open universe, respectively. The corresponding Ricci scalar is
\begin{equation}
R=6\left(\frac{\ddot{a}}{N^{2}a}+\frac{\dot{a}^{2}}{N^{2}a^{2}}-\frac{\dot{a}\dot{N}}{aN^{3}}+\frac{k}{a^{2}}\right).
\end{equation}
Working with the Einstein-Hilbert action
\begin{equation}
S_{EH}=\int dtd^3x\sqrt{-det~g}R[g],
\end{equation}
the Lagrangian for a flat universe takes the form
\begin{equation}
L = \int d^{3}x\left(-\frac{1}{8\pi G}\frac{3\dot a^2 a}{N}\right).
\label{frw-lagrangian}
\end{equation}
where $G$ is Newton's constant. The spatial slice $\Sigma$ is topologically $\mathbb{R}^{3}$, since this space is non-compact the spatial integral in the lagrangian diverges. Restricting the integration to a fiducial cell $\mathcal{V}$ (which we can safely do due to homogeneity), taking a homogeneous scalar field $(\phi=\phi(t))$
we have
\begin{equation}
L = -\frac{V_0}{8\pi G}\frac{3\dot a^2 a}{N} + \frac{\dot{\phi}^{2}a^3}{2N},
\label{frw-lagrangian}
\end{equation}
where $V_{0}=\int_{\mathcal{V}} d^{3}x$ is the coordinate volume of the auxiliary cell $\mathcal{V}$. The momenta $p_{I}= \frac{\partial {L} }{\partial \dot q^{I}}$ are
\begin{equation}
p_a =\frac{\partial {L}}{\partial \dot a} = -\frac{6a\dot a}{N},\quad
p_\phi = \frac{\partial {L}}{\partial \dot \phi}=\frac{a^3\dot{\phi}}{N},
\label{momenta}
\end{equation}
where we have set the fiducial volume equal to one. In the LQC paradigm a great deal of care has been taken in order to ensure that the physical content of the theory do not take into account this fiducial structure.
Taking the Legendre transform, $L_{\mbox{can}}=p_{\mu}\dot{q}^{\mu}-N \mathcal{H}$, where $N$ acts as a Lagrange multiplier which imposes the Hamiltonian constraint $\mathcal{H}=0$, we have
\begin{equation}
H = Na^{-3}\left[-\frac{2\pi G}{3}a^2p_a^2+\frac{p_\phi^2}{2}\right]. \label{hamiltonian}
\end{equation}
With the canonical relations
\begin{equation}
\{q^{I},p_{J}\}=\delta^{I}_{J},
\end{equation}
and the Hamiltonian at hand, Hamilton equations
\begin{equation}
\dot{p}_{I}=\{p_{I},H\}=-\frac{\partial H}{\partial q^{I}};\quad\dot{q}^{I}=\{q^{I},H\}=\frac{\partial H}{\partial p_{I}},
\label{hamiltoneqs}
\end{equation}
read
\begin{equation}
\dot{p}_{a}=-\frac{2\pi G p^{2}_{a}}{3a^2}+\frac{3p^2_\phi}{2a^4}\label{piaeq},\quad\dot{p}_{\phi}=0\label{pifieq},\quad\dot{a}=-\frac{4\pi G p_{a}}{3a}\label{aeq},\quad\dot{\phi}=\frac{p_{\phi}}{a^3},
\label{fieq}
\end{equation}
where $N=1$ has been set. Taking into account the field equations and the Hamiltonian constraint one arrives at the Friedmann equation
\begin{equation}
\mathrm{H}^{2}=\frac{8\pi G}{3}\rho,\label{friedmann_ordinary}
\end{equation}
$\mathrm{H}=\frac{\dot{a}}{a}$ is called the Hubble parameter, and, in this particular case of a free standard homogeneous scalar field, $\rho=\frac{\dot{\phi}^{2}}{2}=\frac{p^{2}_{\phi}}{2V^{2}}$ ($V=a^3$) is the energy density.

\section{Connection Variables and Effective Dynamics}\vspace{5mm}

\subsection{The FLRW Model in the Ashtekar-Barbero Variables}\hfill \break
In this section, we recall the formulation of the open FLRW model in the Ashtekar-Barbero variables, with a free massless scalar field.  The Ashtekar-Barbero variables cast General Relativity in the form of a gauge theory, in which phase space is described by an $\frak{su}(2)$ gauge connection, the Ashtekar-Barbero connection $A^{i}_{a}$, and its canonical conjugate momentum, the densitized triad $E^{a}_{i}$. In both, the Ashtekar-Barbero and the densitized triad, $i,j, \dots ,$ denote internal $\frak{su}(2)$ indices while $a,b,\dots,$ denote spatial indices. These quantities are defined as
\begin{equation}
A^{i}_{a}=\Gamma^{i}_{a}+\gamma K^{i}_{a},\quad E^{a}_{i}=\sqrt{q}e^{a}_{i},
\end{equation}
where $K^{i}_{a}=K_{ab}e^{a}_{j}$, with $e^{i}_{a}$ and $e^{a}_{i}$ the triad and co-triad, respectively, which satisfied $e^{i}_{a}e^{a}_{j}=\delta^{i}_{j}$, and $q_{ab}=\delta_{ij}e^{i}_{a}e^{j}_{b}$; $\Gamma^{i}_{a}$ is defined through the spin connection $\omega^{i}_{aj}$ compatible with the triad  by the relation $\omega^{i}_{aj}=\epsilon_{ijk}\Gamma^{i}_{a}$ with $\epsilon_{ijk}$ the totally antisymmetric symbol and $\nabla_{a}e^{i}_{b}+\omega^{i}_{aj}e^{j}_{b}=0$ ($\nabla_{a}$ being the usual spatial covariant derivative); $\gamma$ a real constant called the Barbero-Immirzi parameter. The canonical pair has the following Poisson structure
\begin{equation}
\{A^{a}_{i}(x),E^{j}_{b}(y)\}=8\pi G\gamma\delta^{j}_{i}\delta^{a}_{b}\delta(x,y),
\end{equation}
with $\delta(x,y)$ the Dirac delta distribution on the space-like hypersurface $\Sigma$.
In these variables, the gravitational Hamiltonian density takes the form
\begin{equation}
C_{grav}=\left(\frac{1}{|E|}\epsilon_{ijk}\left[F^{i}_{ab}-(1+\gamma^{2})\epsilon^{imn}K^{m}_{a}K^{n}_{b}\right]E^{aj}E^{bk}\right), \label{C_grav_density}
\end{equation}
where $F^{i}_{ab}=\partial_{a}A^{i}_{b}-\partial_{b}A^{i}_{a}+\epsilon_{ijk}A^{j}_{a}A^{k}_{b}$ is the curvature of the Ashtekar-Barbero connection and $E$ the determinant of the densitized triad. 

From the gravitational Hamiltonian density (\ref{C_grav_density}), one can cast the gravitational Hamiltonian through
\begin{equation}
H_{grav}=\int d^{3}x NC_{grav},
\end{equation}
in the case for spatially flat homogeneous models reduces to \cite{ashtekar-lqc}
\begin{equation}
H_{grav}=-\gamma^{2}N\int_{\mathcal{V}}d^{3}x\frac{1}{|E|}\epsilon_{ijk}F^{i}_{ab}E^{aj}E^{bk}, \label{h_grav}
\end{equation}
where the integral is taken in a fiducial cell $\mathcal{V}$. When imposing also isotropy, the connection and triad can be described by  parameters $c$ and $p$, respectively, defined by \cite{ashtekar-lqc}
\begin{equation}
A^{i}_{a}=c V_{0}\, ^{o}e^{i}_{a},\quad E^{i}_{a}=pV_{0}\sqrt{^{o}q}\, ^{o}e^{a}_{i},
\end{equation}
where $^{o}q_{ab}$ is a fiducial flat metric
and $^{o}e^{a}_{i}$, $^{o}e^{i}_{a}$ are constant triad and co-triad compatible with $^{o}q_{ab}$; $V_{0}$ is the volume of $\mathcal{V}$ with respect to $^{o}q_{ab}$. These variables do not depend on the choice of the fiducial metric.
The relation among these variables and the usual geometrodynamical variables is
\begin{eqnarray}
c=V_{0}^{1/3}\gamma\dot{a},\quad p=V_{0}^{2/3}a^{2}, 
\end{eqnarray}
where the canonical relations for this last two variables obey
\begin{equation}
\{c,p\}=\frac{8\pi G\gamma}{3}.
\end{equation}
Performing the following change of variables
\begin{equation}
\beta=\frac{c}{\sqrt{p}},\quad V=p^{\frac{3}{2}},
\end{equation} 
the Hamiltonian (\ref{h_grav}), with a free massless scalar field as the matter content, takes the form
\begin{equation}
H=N\left(-\frac{3}{8\pi G\gamma^{2}}\beta^{2}V+\frac{p^{2}_{\phi}}{2V}\right),
\label{ham1}
\end{equation}
with the canonical relations for $\beta$ and $V$ being
\begin{equation}
\{\beta,V\}=4\pi G\gamma.
\end{equation}
The holonomy correction due to Quantum Gravity effects is coded in the replacement \cite{ashtekar-lqc}
\begin{equation}
\beta\mapsto\frac{\sin({\lambda\beta})}{\lambda},
\end{equation}
where $\lambda^{2}=4\sqrt{3}\pi\gamma\ell^{2}_{p}$ is the smallest eigenvalue of the area operator in the full Loop Quantum Gravity \cite{thiemann}.
The resulting effective Hamiltonian (taking $N=1$) is thus given by
\begin{equation}
H_{eff}=-\frac{3}{8\pi G\gamma^{2}\lambda^{2}}\sin^{2}(\lambda\beta)V+\frac{p^{2}_{\phi}}{2V}, \
\label{ham2}
\end{equation}
which in turn, leads us to obtain the following field equations
\numparts
\begin{eqnarray}\label{beta_conm}
\dot{\beta}=4\pi G\gamma\frac{\partial H_{eff}}{\partial V}=-\frac{3}{\gamma\lambda^{2}}\sin^{2}(\lambda\beta)-4\pi G\gamma\frac{p^2_{\phi}}{2V^{2}},\\
\dot{V}=-4\pi G\gamma\frac{\partial H_{eff}}{\partial\beta}=\frac{3}{\gamma\lambda}V\sin(\lambda\beta)\cos(\lambda\beta)\label{V_conm},\\
\dot{\phi}=\frac{\partial H_{eff}}{\partial p_{\phi}}=\frac{p_{\phi}}{V},\label{fi_conm}\\
\dot{p}_{\phi}=-\frac{\partial H_{eff}}{\partial\phi}=0.\label{pfi_conm}
\end{eqnarray}
\endnumparts
Since $\frac{\dot{V}}{3V}=\frac{\dot{a}}{a}=\mathrm{H}$, where $\mathrm{H}$ is the Hubble parameter; then, taking into account the effective Hamiltonian constraint, $H_{eff}\approx 0$, and the field equation for $V$, we have

\begin{equation}
\mathrm{H}^{2} =\left(\frac{\dot{V}}{3V}\right)^{2}=\frac{8\pi G}{3}\rho\left(1-\frac{\rho}{\rho_{max}}\right),
\label{eff-FRW}
\end{equation}
where $\rho=\frac{(\dot\phi)^2}{2}=\frac{p^{2}_{\phi}}{2V^{2}}=\frac{3}{8\pi G\gamma^{2}\lambda^{2}}\sin^{2}(\lambda\beta)$ and $\rho_{max}$ is the maximum value that $\rho$ can take in view of the effective Hamiltonian constraint, that is, $\rho_{max}=\frac{3}{8\pi G\gamma^{2}\lambda^{2}}$. The turning points of the volume function occur at $\beta=\pm\frac{\pi}{2\lambda}$, which correspond to a bounce. Equation (\ref{eff-FRW}) is the modified Friedmann equation, which incorporates holonomy corrections due to Loop Quantum Gravity. In the limit $\lambda\rightarrow0$ (no area gap) we recover the  ordinary Friedmann equation, equation (\ref{friedmann_ordinary}).

The relational evolution of $V$ in terms of $\phi$ is given by
\begin{equation}
\frac{dV}{d\phi}=\frac{dV}{dt}\frac{dt}{d\phi}=\frac{3}{\gamma\lambda}\sin(\lambda\beta)\cos(\lambda\beta)\frac{V}{p_{\phi}}=\sqrt{12\pi G}V\left(1-\frac{\rho}{\rho_{max}}\right)^{1/2}
\label{relational-dy}
\end{equation}
where we have used the field equations for $V$ and $\phi$, and the effective Hamiltonian constraint.

\section{Deformations in Classical and Quantum Mechanics}\vspace{5mm}

The ideas of deformed minisuperspace, in connection to noncommutative cosmology, were introduce in the seminal work of Garcia-Compean, et al \cite{hugo}. The introduction of a deformation to the minisuperspace, in order to incorporate an effective noncommutativity, made it possible to avoid working with a noncommutative theory of gravity, and still get relavant physics of the Kantoswki-Sachs cosmological model.
In the canonical quantum cosmological scenario, the Wheeler-deWitt equation is responsible for the description of the quantum behavior of the Universe. An alternative approach to study quantum mechanical effects, is to introduce deformations to the Poisson structure of classical phase space; the approach is an equivalent path to quantization and is part of a complete and consistent type of quantization known as deformation quantization \cite{flato}. In \cite{Cordero}, the authors apply the deformation quantization formalism to cosmological models in the flat minisuperspace. Within this setup, relevant minisuperspace cosmological models are studied in detail, namely the de Sitter and Katowski-Sachs, among others. 

In the deformation phase space approach, the deformation is introduced by the Moyal brackets $\{f,g\}_{\alpha}=f\star_{\alpha}g-g\star_{\alpha}f$, where the product between functions is replaced by the Moyal product
\begin{equation}
(f\star g)(x)=\exp\left[\frac{1}{2}\alpha^{ab}\partial_a^{(1)}\partial_b^{(2)}\right]f(x_1)g(x_2)\vert_{x_1=x_2=x},
\end{equation}
such that
\begin{eqnarray}
\alpha =
\left( {\begin{array}{cc}
 \theta_{ij} & \delta_{ij}+\sigma_{ij}  \\
- \delta_{ij}-\sigma_{ij} & \beta_{ij}  \\
 \end{array} } \right),
\end{eqnarray}
where the $2\times 2$ matrices $\theta_{ij}$ and $\beta_{ij}$ are assume to be antisymmetric and represent the noncommutativity in the coordinates and the momenta, respectively. The resulting $\alpha$-deformed algebra for the phase space is
\begin{equation}
\{x_i,x_j\}_{\alpha}=\theta_{ij}, \;\{x_i,p_j\}_{\alpha}=\delta_{ij}+\sigma_{ij},\; \{p_i,p_j\}_{\alpha}=\beta_{ij}.\label{alg}
\end{equation}
An alternative to derive an algebra similar to (\ref{alg}), is making the following transformation on the classical phase variables $\{x,y,P_x,P_y\}$
\numparts
\begin{eqnarray}
\hat{x}=x+\frac{\theta}{2}P_{y}, \qquad \hat{y}=y-\frac{\theta}{2}P_{x},\label{nctrans1}\\
\hat{P}_{x}=P_{x}-\frac{\beta}{2}y, \qquad \hat{P}_{y}=P_{y}+\frac{\beta}{2}x, \label{nctrans2}
\end{eqnarray}
\endnumparts
where particular expressions for the deformations have been consider, namely $\theta_{ij}=-\theta\epsilon_{ij}$ and $\beta_{ij}=\beta\epsilon_{ij}$. The resulting algebra is the same as (\ref{alg}), but the Poisson brackets are different in the two algebras. For equations (\ref{alg}), the brackets are the $\alpha$-deformed ones and are related to the Moyal product; for the other algebra the brackets are the usual Poisson brackets, which read
\begin{equation}
\{\hat{y},\hat{x}\}=\theta,\; \{\hat{x},\hat{P}_{x}\}=\{\hat{y},\hat{P}_{y}\}=1+\sigma,\; \{\hat{P}_y,\hat{P}_x\}=\beta,\label{dpa}
\end{equation}
where $\sigma=\theta\beta/4$.

In this two scenarios the outcome can be summarize as follows: first, in the deformation quantization formalism, the $\alpha$-deformed algebra (\ref{alg}), when applied to a given system characterised by a canonical Hamiltonian $H$, one gets a deformed Hamiltonian $H^{\prime}$ given rise to deformed equations of motion. Secondly, constructing a deformed phase space introducing the transformation (\ref{nctrans1}) and (\ref{nctrans2}), it is possible to formulate a Hamiltonian, formally analogous to a canonical one, but with the variables that obey the modified algebra (\ref{dpa}). In this paper we will work under the lines of thought of the latter.

\section{Noncommutativity in FLRW Loop Quantum Cosmology.}\vspace{5mm}

\subsection{FLRW without LQC corrections.}\hfill\break
We will start this section analysing the FLRW model for a free scalar field, and without the LQC corrections. In Ashtekar variables, and volume representation, the classical Hamiltonian takes the for 
\begin{equation}
H=-\frac{3}{8\pi G\gamma^{2}}\beta^{2}V+\frac{p^{2}_{\phi}}{2V},
\label{ham1}
\end{equation}
where $N=1$, has been taken. In view of the above discussion, we would like to consider effects of a deformed algebra (noncommutativity)
\begin{equation}
\{\beta^{nc},\phi^{nc}\}=\theta,\quad\{\beta^{nc},V^{nc}\}=4\pi G\gamma,\quad\{\phi^{nc},p^{nc}_{\phi}\}=1,
\end{equation}
with the remaining brackets zero.
The above relations can be implemented by working with the shifted variables
\begin{equation}
\beta^{nc}=\beta+a\theta p_{\phi},\quad\phi^{nc}=\phi+b\theta V,\quad V^{nc}=V,\quad p^{nc}_{\phi}=p_{\phi},
\label{nc-rel}
\end{equation}
where $a$ and $b$ satisfy the relation $4\pi G\gamma b-a=1$.

The deformed Hamiltonian constraint arising from the steps above has the same functional form as (\ref{ham1}), but is valued in the new noncommuting variables \cite{barbosa}, that is
\begin{equation}
H^{nc}=-\frac{3}{8\pi G\gamma^{2}}(\beta^{nc})^{2}V+\frac{p^{2}_{\phi}}{2V},
\label{hamnc}
\end{equation}
in the limit $\theta\rightarrow0$ this Hamiltonian reduces to the usual one (\ref{ham1}). The noncommutative equations of motion are
\numparts
\begin{eqnarray}\label{system1}
\dot{\beta}=4\pi G\gamma\frac{\partial H^{nc}}{\partial V}=-\frac{3}{2\gamma}(\beta^{nc})^{2}-4\pi G\gamma\frac{p^{2}_{\phi}}{2V^{2}},\\
\dot{V}=-4\pi G\gamma\frac{\partial H^{nc}}{\partial\beta}=\frac{3}{\gamma}V\beta^{nc},\\
\dot{\phi}=\frac{\partial H^{nc}}{\partial p_{\phi}}=-\frac{3a\theta V}{4\pi G\gamma^{2}}\beta^{nc}+\frac{p_{\phi}}{V},\\
\dot{p}_{\phi}=-\frac{\partial H^{nc}}{\partial\phi}=0,\label{system44}
\end{eqnarray}
\endnumparts
as expected, these Hamiltonian equations reduce to the usual ones when taking $\theta\rightarrow0$. Furthermore, the energy density is given by
\begin{equation}
\rho=\frac{(\dot\phi)^{2}}{2}=\frac{1}{2V}\left(-\frac{3a\theta V}{4\pi G\gamma^{2}}\beta^{nc}+\frac{p_{\phi}}{V}\right)^{2}.
\end{equation}
The exact solutions to the system of equations (\ref{system1}-\ref{system44}) are
\numparts
\begin{eqnarray}
\beta=\frac{\gamma-3aB\theta t+aAB\gamma\theta}{3t-A\gamma},\\
V=C(3t-B\gamma),\\
\phi=-\frac{3aC\theta t}{4\pi G\gamma}+\frac{B\log(3Ct-AC\gamma)}{3C}+D,\\
p_{\phi}=B,
\end{eqnarray}
\endnumparts
where $A,B,C,D$ are integration constants. These solutions reduce to the usual ones when taking $\theta\rightarrow0$. We note that the solution for the volume function is the same as in the usual commutative case.

\subsection{FLRW with LQC corrections.}\hfill\break
Now we want to implement noncommutativity by working in the shifted variables (\ref{nc-rel}) defined atop. Since we are not actually performing a deformation of the symplectic structure, the loop quantization of the original variables can be carried as usual.\\

Considering the steps above, and taking into account the ideas posed in the last section, we can therefore implement the effects of relation (\ref{nc-rel}) at the effective scheme of LQC by considering the Hamiltonian
\begin{equation}
H^{nc}_{eff}=-\frac{3}{8\pi G\gamma^{2}\lambda^{2}}\sin^{2}(\lambda\beta^{nc})V+\frac{p^{2}_{\phi}}{2V},
\label{nc-frw-ham}
\end{equation}
where the noncommutative effective field equations result to be
\numparts
\begin{eqnarray}
\dot{\beta}=4\pi G\gamma\frac{\partial H^{nc}_{eff}}{\partial V}=-\frac{3}{2\gamma\lambda^{2}}\sin^{2}(\lambda\beta^{nc})-4\pi G\gamma\frac{p^{2}_{\phi}}{2V^{2}},\label{beta_nc}\\
\dot{V}=-4\pi G\gamma\frac{\partial H^{nc}_{eff}}{\partial\beta}=\frac{3}{\gamma\lambda}V\sin(\lambda\beta^{nc})\cos(\lambda\beta^{nc})\label{V_nc},\\
\dot{\phi}=\frac{\partial H^{nc}_{eff}}{\partial p_{\phi}}=-\frac{3a\theta V}{4\pi G\gamma^{2}\lambda}\sin(\lambda\beta^{nc})\cos(\lambda\beta^{nc})+\frac{p_{\phi}}{V},\label{fi_nc}\\
\dot{p}_{\phi}=-\frac{\partial H^{nc}_{eff}}{\partial\phi}=0\label{pfi_nc},
\end{eqnarray}
\endnumparts
again, in the limit $\theta\rightarrow0$ we recover the commutative field equations.
Due to the field equation for $\phi$, we note that now the matter density $\rho=\frac{\dot{\phi}^{2}}{2}$ is not given only by $\frac{p^{2}_{\phi}}{2V^{2}}$, but by
\begin{equation}
\rho=\frac{1}{2}\left(-\frac{3a\theta}{4\pi G\gamma^{2}\lambda}\sin(\lambda\beta^{nc})\cos(\lambda\beta^{nc})+\frac{p_{\phi}}{V}\right)^{2}.
\end{equation}
In order to construct the analog of the Friedmann equation we would need to obtain a relation for $p_\phi$ in terms of $\dot{\phi}$, but, due to the equation for $\dot{\phi}$ being now more tangled, such relation can not be obtained.The relational evolution of $V$ in terms of $\phi$ is now given by
\begin{equation}\fl
\left(\frac{dV}{d\phi}\right)^{2}=\left(\frac{3V}{\gamma\lambda}\right)^{2}\sin^{2}(\lambda\beta^{nc})\cos^{2}(\lambda\beta^{nc})\left[\frac{3\theta}{8\pi G\gamma^{2}\lambda^{2}}\sin^{2}(\lambda\beta^{nc})\cos^{2}(\lambda\beta^{nc})+\frac{p_{\phi}}{V}\right]^{-2},\label{nc-relational-dy}
\end{equation}
when taking $\theta\rightarrow0$ this relation reduces to (\ref{relational-dy}).

We note that the equations for $\beta(t)$ and $V(t)$ are very similar to their commutative counterparts; the equation for $\phi(t)$ is the only one being considerably more tangled, as pointed out above. The equation for $p_{\phi}(t)$ is exactly the same, indicating that $p_\phi$ is a constant of motion.

We observe that $\dot{V}=0$ for $\beta=\pm\frac{\pi}{2\lambda}-a\theta p_\phi$, furthermore, $\ddot{V}\vert_{\beta=\pm\frac{\pi}{2\lambda}-a\theta p_{\phi}}=\frac{6V}{\gamma^{2}\lambda^{2}}>0$. So these values of $\beta$ correspond to a minimum in the volume function and to a change in sign in $\dot{V}$. Indicating that a bounce takes place. The energy density at the bounce is $\rho\vert_{\beta=\pm\frac{\pi}{2\lambda}-a\theta p_{\phi}}=\frac{3}{8\pi G\gamma^{2}\lambda^{2}}$, the same as in the commutative case. This means that the key features of LQC are mantained.

The equation for $\dot{\beta}$ can be solved at once with the help of the Hamiltonian constraint ($H_{eff}^{nc}\approx 0$):
\begin{equation}
\beta(t)=\frac{1}{\lambda}\mathrm{arccot}\left(\frac{3t+2A\gamma\lambda^2}{\gamma\lambda}\right)-a\theta p_{\phi},
\label{sol-beta}
\end{equation}
$A$ being an integration constant. If the bounce take place at $t=0$ (like in standard LQC) then we must have $A=0$ and $\beta(t)$ takes the  form
\begin{equation}
\beta(t)=\frac{1}{\lambda}\mathrm{arccot}\left(\frac{3t}{\gamma\lambda}\right)-a\theta p_{\phi},
\end{equation}
this solution reduces to the commutative one \cite{sols} upon taking $\theta\rightarrow 0$.
\begin{center}
	\includegraphics[scale=0.50]{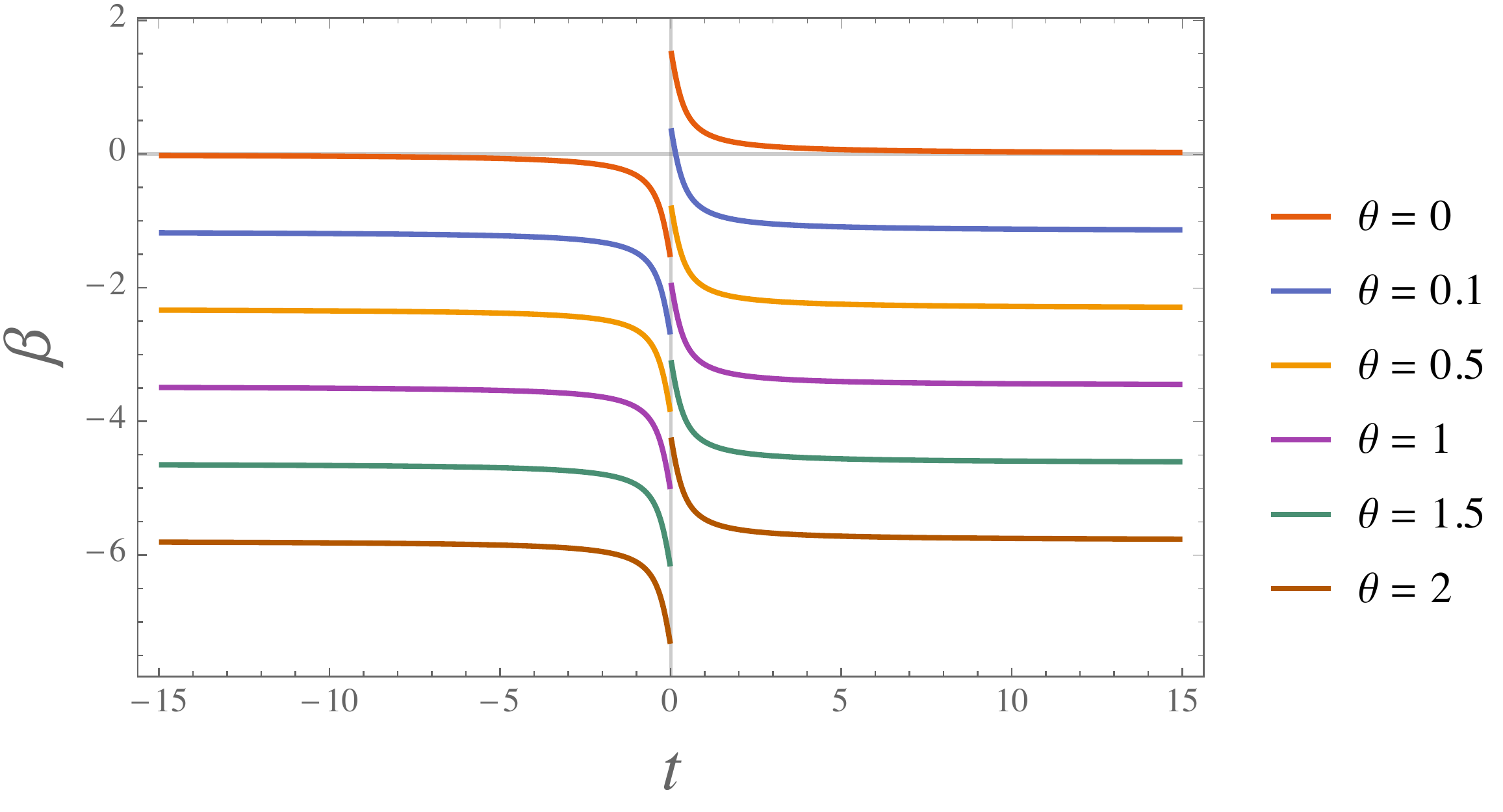}
	\captionof{figure}{The behavior of $\beta(t)$ for different values for the noncommutative parameter, with the remaining constants fixed.}
\end{center}
By substituting the solution for $\dot{\beta}$ in the equation for $\dot{V}$ we obtain
\begin{equation}
V(t)=B\sqrt{\gamma^{2}\lambda^{2}+9t^{2}},
\label{sol-v}
\end{equation}
$B$ being an integration constant. This solution coincides with the one obtained in the commutative case \cite{sols}. Therefore, the behavior of the volume function in this noncommutative construct is the same as in the usual LQC.

\begin{center}
	\includegraphics[scale=0.50]{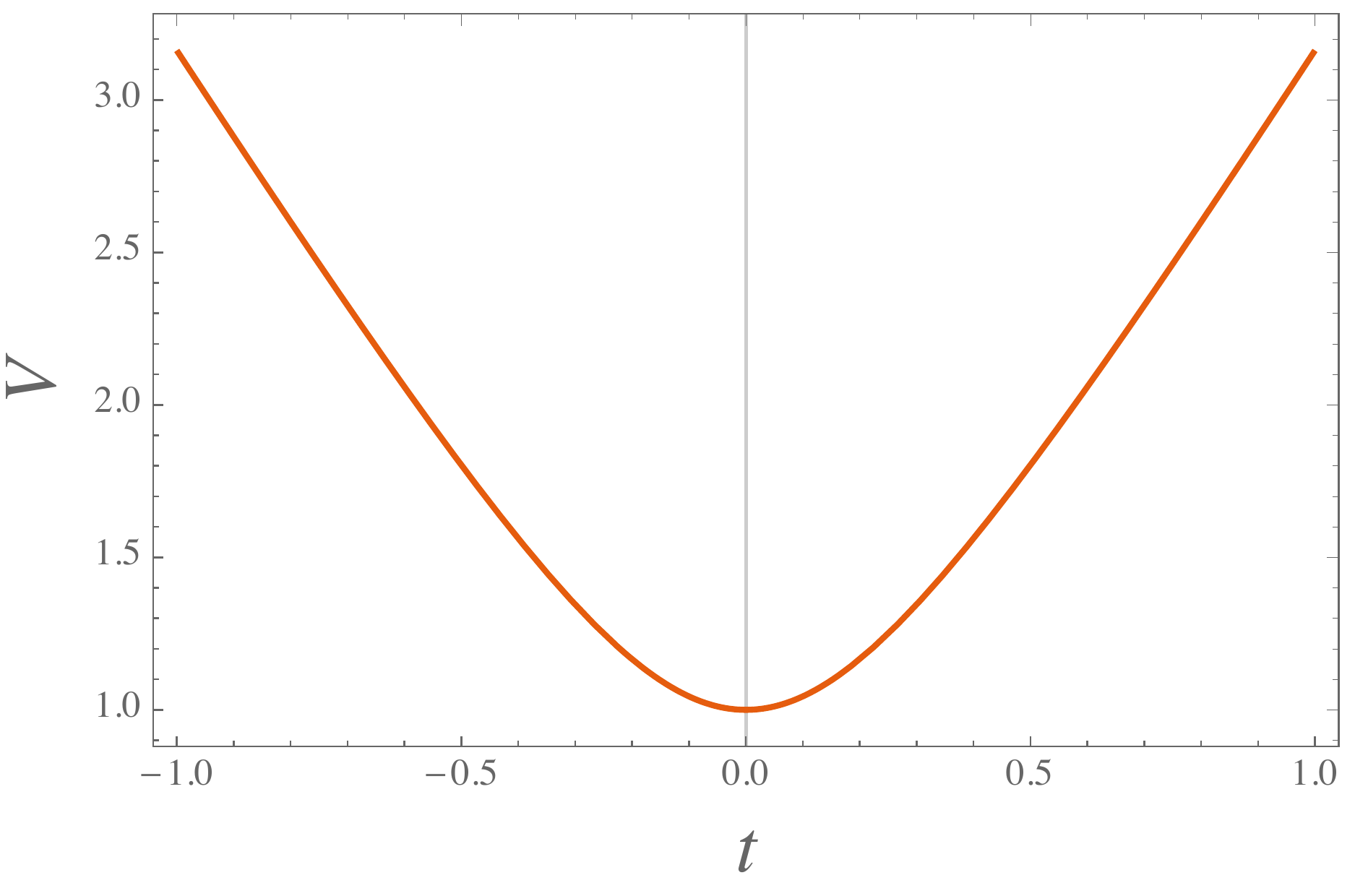}
	\captionof{figure}{The behavior of the volume function is the same as in standard LQC.}
\end{center}


\begin{center}
	\includegraphics[scale=0.50]{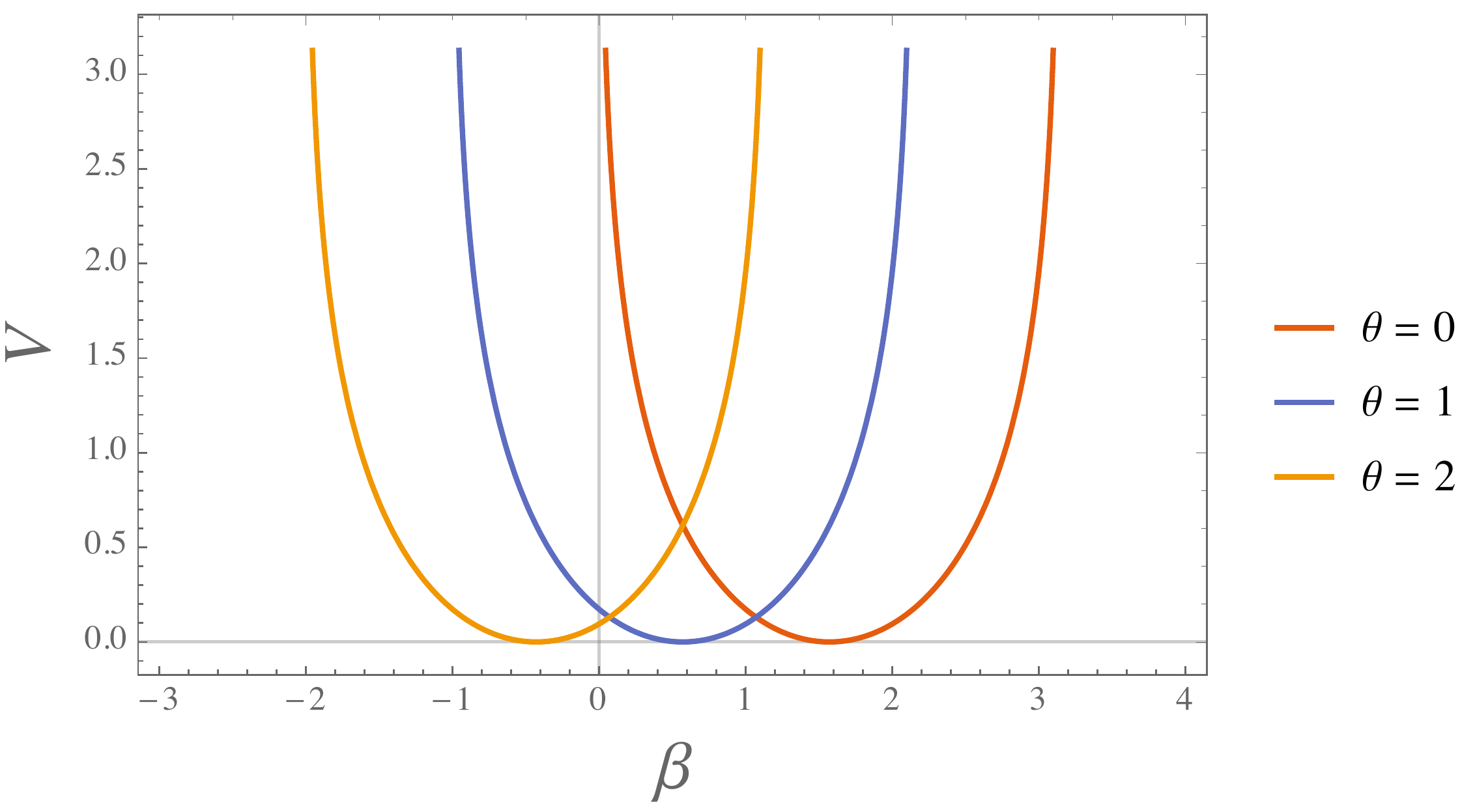}
	\captionof{figure}{The behavior of the volume as a function of $\beta$ for different values of the noncommutative parameter, with the remaining constants fixed. We observe that the bounce can be shifted by tunning $\theta$.}
\end{center}

Substituting the solutions for $\beta(t)$ and $V(t)$ in the equation for $\dot{\phi}$ we get
\begin{equation}
\phi(t)=C+\frac{p_{\phi}}{3B}\mathrm{log}\left(3t+\sqrt{\gamma^2\lambda^2+9t^2}\right)-\frac{aB\theta\sqrt{\gamma^2\lambda^2+9t^2}}{4\pi G\gamma},
\label{sol-fi}
\end{equation}
where $C$ is another integration constant. This solution also reduces to the commutative one upon taking $\theta\rightarrow 0$ \cite{sols}.

Employing the solution for $\phi(t)$ to construct the energy density $\rho=\frac{\dot{\phi}^{2}}{2}$, we observe that its behavior is similar to the one encountered in LQC. We note that this energy density overlaps very fast to the one found in LQC, as we take smaller values for the noncommutative parameter $\theta$.
\begin{center}
	\includegraphics[scale=0.50]{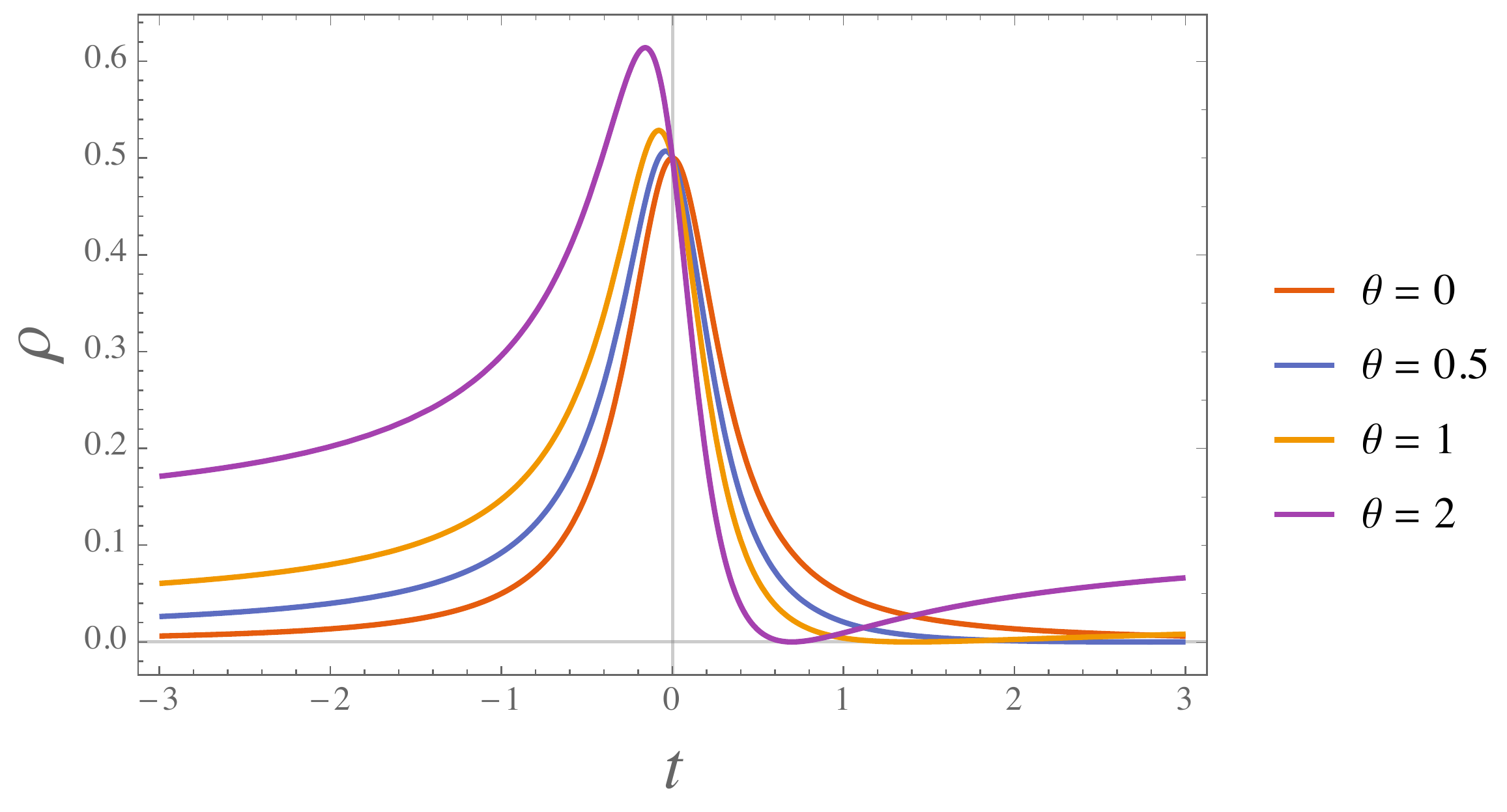}
	\captionof{figure}{The behavior of the energy density for different vlaues of $\theta$ with the remaining constants fixed. We observe that the form of the energy density is not retained.}
\end{center}
\subsection{Noncommutativity in Momentum Sector of FLRW}\hfill\break
In this section we would like to study the effects of noncommutativity in the momentum sector. The way to proceed is in the same manner as the past sections. Consider a deformed algebra
\begin{equation}
\{V^{nc},p^{nc}_{\phi}\}=\theta,\quad\{\beta^{nc},V^{nc}\}=4\pi G\gamma,\quad\{\phi^{nc},p^{nc}_{\phi}\}=1,
\label{ncm-free}
\end{equation}
with the remaining brackets being zero.
The above relations can be implemented working with the shifted variables
\begin{equation}
V^{nc}=v+a\theta\phi,\quad p_{\phi}^{nc}=p_\phi+b\theta\beta,\quad \beta^{nc}=\beta,\quad\phi^{nc}=\phi,
\label{nc-rel2}
\end{equation}
where $a$ and $b$ satisfy the relation $a-4\pi G\gamma b=1$.

The deformed Hamiltonian takes the form
\begin{equation}
H^{nc}_{eff}=-\frac{3}{8\pi G\gamma^{2}\lambda^{2}}\sin^{2}(\lambda\beta)V^{nc}+\frac{\left(p_{\phi}^{nc}\right)^{2}}{2 V^{nc}},
\label{nc2-frw-ham}
\end{equation}
and the noncommutative effective field equations are
\numparts
\begin{eqnarray}
\dot{\beta}=4\pi G\gamma\frac{\partial H^{nc}_{eff}}{\partial V}=-\frac{3}{2\gamma\lambda^{2}}\sin^{2}(\lambda\beta),\\
\dot{V}=-4\pi G\gamma\frac{\partial H^{nc}_{eff}}{\partial\beta}=\frac{3}{\gamma\lambda}V^{nc}\sin(\lambda\beta)\cos(\lambda\beta)-\frac{4\pi G\gamma b\theta p_{\phi}^{nc}}{V^{nc}},\\
\dot{\phi}=\frac{\partial H^{nc}_{eff}}{\partial p_{\phi}}=\frac{p_{\phi}^{nc}}{V^{nc}},\\
\dot{p}_{\phi}=-\frac{\partial H^{nc}_{eff}}{\partial\phi}=\frac{3a\theta}{8\pi G\gamma^{2}\lambda^{2}}\sin^{2}(\lambda\beta)+a\theta\frac{\left(p_{\phi}^{nc}\right)^{2}}{2\left(V^{nc}\right)^{2}},
\end{eqnarray}
\endnumparts
in the limit $\theta\rightarrow0$ we recover the commutative field equations.

The equation for $\beta(t)$ is the same as in standard Loop Quantum Cosmology, the remaining equations of motion are now more complex and an exact solution could not be found. 

With the help of the Hamiltonian constraint, we observe that $\rho=\frac{3}{8\pi G\gamma^{2}\lambda^{2}}\sin^{2}(\lambda\beta)$, the same relation as in standard Loop Quantum Cosmology, and hence, the maximum value that $\rho$ can attain is the same as in the commutative case, $\rho_{max}=\frac{3}{8\pi G\gamma^{2}\lambda^{2}}$. This maximum value is reached, according to the equation for $\beta(t)$, at $t=0$, as in the commutative case. This indicate that, when implementing the noncommutativity defined by (\ref{ncm-free}), not only a bounce ocurs, but that this bouce has the same characteristics as the one in standard LQC, since the critical density asociated with it is the same and the time when it happens is the same.
 
\section{Discusion and Final Remarks}\hfill \break
A simple noncommutative extension of the open FLRW model in Loop Quantum Cosmology has been constructed, through the introduction of a deformation at the effective scheme of Loop Quantum Cosmology. These models could incorporate effective corrections from both Loop Quantum Gravity and Noncommutative Geometry.

When introducing noncommutativity in the configuration sector, it is observed from the equation for $\dot{V}$ that a bounce occurs when $\beta=\frac{\pi}{2\lambda}-a\theta p_\phi$, therefore, the bounce can be shifted in time by tunning $\theta$; morover, at this value of $\beta$, the noncommutative density is the same as in Effective Loop Quantum Comology, $\rho_{max}$. Even when the behavior of the energy density is similar to the one of standard Loop Quantum Cosmology, as we consider smaller and smaller values for the noncommutative parameter, the form of the density function is not retained. 

For the case of noncommutativity in the momentum sector, it was observed that the equation for $\beta(t)$ agrees with the one in standard Loop Quantum Cosmology. The solutions for the remaining degrees of freedom could not be obtained analytically. It is observed that the behavior of the energy density is the same as in the commutative case: the characteristics of the bounce match those of standard LQC. This signals more compatibility between this kind of noncommutativity and the LQC pardigm than the one in the configuration sector. 

Finally, we conclude that a deformation in the momentum sector of the phase space spanned by the flat FLRW model with a standard free scalar field is more compatible with the LQC paradigm than a deformation in the configuration sector. Of course, further research is required to establish how deep this compatibility is. Some of this additional analysis will be reported by the authors elsewhere. 
\hfill\break

\ack{This work was partially supported by CONACyT  167335, 179881 grants. PROMEP grants UGTO-CA-3. 
A.E.G. was supported by a CONACYT graduate fellowship during a major part of the realization of this work. S.P.P. was partially supported by SNI-CONACyT.
This work is part of the collaboration within the Instituto Avanzado de Cosmolog\'{\i}a and Red PROMEP: Gravitation
and Mathematical Physics under project {\it Quantum aspects of gravity in cosmological models, phenomenology and geometry of
space-time}.
A.E.G. wishes to thank J. Socorro, M. Sabido, Alejandro Corichi, J. Daniel-Reyes and Héctor H. Hernández for discussions.}\\

	\end{document}